\documentclass[aps,prb,twocolumn, superscriptaddress]{revtex4-2}

\usepackage{graphicx}

\usepackage{multirow}
\usepackage{textgreek}
\usepackage{babel}
\usepackage{amsmath,amssymb}

\usepackage{color} 

\begin{document}

\title{Revealing tilt-driven structural heterogeneity in hybrid improper ferroelectrics by spatially-resolved crystallography}

\author{Evie Ladbrook} 
\affiliation{Department of Chemistry, University of Warwick, Gibbet Hill, Coventry, CV4 7AL, United Kingdom}

\author{Jon P. Wright} 
\affiliation{European Synchrotron Radiation Facility, 38000 Grenoble, France}

\author{Mark S. Senn} 
\email{m.senn@warwick.ac.uk}
\affiliation{Department of Chemistry, University of Warwick, Gibbet Hill, Coventry, CV4 7AL, United Kingdom}

\date{\today}%

\begin{abstract}

Domain walls in ferroic oxides can host emergent functionalities that differ fundamentally from those of the surrounding bulk, making their microscopic structure central to understanding and controlling macroscopic behaviour. We use scanning three-dimensional X-ray diffraction to map the spatial evolution of coupled octahedral distortion modes across ferroelastic domain walls in the hybrid improper ferroelectric Ca$_{2.15}$Sr$_{0.85}$Ti$_{2}$O$_{7}$. By combining tomographic reconstruction with symmetry-adapted analysis of the octahedral rotation and tilt and associated strain modes, we map their spatial evolution across individual walls in the bulk crystal. We find that the walls are intrinsically extended, with widths far exceeding those expected for conventional Ising-like ferroic interfaces. The wall structure is characterised by a continuous rotation of the tilt order parameter through a tetragonal intermediate. These results provide new microscopic insight into the structural mechanisms governing ferroelastic wall formation, ferroelectric switching, and emergent functionality at domain walls.

\end{abstract}

\maketitle

\section{Introduction}

Spatial variations in structure, chemistry, and electronic order play a central role in determining the functional behaviour of complex oxides. In many cases, macroscopic properties are governed not only by the magnitude of an order parameter, but by how it varies in space. In ferroic systems, the spontaneous emergence of an order parameter leads to the formation of domains which are regions of equivalent symmetry distinguished by the orientation of that order parameter. These domains form to minimise the total free energy and are separated by domain walls, across which the order parameter varies over a finite length scale. The configuration and dynamics of domains and domain walls strongly influence macroscopic behaviour, for example, switching under applied fields proceeds via domain wall motion. Domain walls can also exhibit properties distinct from the bulk, including enhanced conductivity in ferroelectrics \cite{seidel_conduction_2009}, superconductivity in ferroelastics \cite{aird_sheet_1998}, and emergent polarity in otherwise non-polar systems \cite{goncalves-ferreira_ferrielectric_2008}. As a result, there is growing interest in domain walls as functional elements in their own right \cite{catalan_domain_2012, nataf_domain-wall_2020, mundy_functional_2017}.

Ferroelectric and ferroelastic walls are often described as Ising-like, where the order parameter is suppressed to zero at the wall centre, resulting in narrow interfaces (typically less than 2 nm for neutral ferroelectric walls \cite{tagantsev_domains_2010}). However, increasing experimental and theoretical evidence points to the existence of wider, structurally coherent walls, in which the order parameter evolves through rotation rather than suppression. For example, Landau theory predicts domain walls tens of nanometres wide in BaTiO$_3$ \cite{yudin_anomalously_2015}, while first-principles calculations reveal mixed Bloch–N\'eel–Ising character in PbTiO$_3$ \cite{lee_mixed_2009}. 

This behaviour is particularly rich in systems with multiple coupled order parameters. In improper ferroelectrics, where the polarisation emerges as a secondary effect to one or more primary order parameters, multiple distinct combinations of the primary distortions can result in domains that are equivalent in polarisation direction but differ structurally. As a result, domain walls can separate regions that differ not only in polarisation but also in other underlying distortions. This coupling between order parameters enables the formation of complex topological textures, including domain wall intersections, loops and vortices. A striking example is the `cloverleaf' vortex structure in YMnO$_3$, where six ferroelectric domains converge at a single point \cite{choi_insulating_2010}.

\begin{figure}[h]
\includegraphics[width=0.48\textwidth]{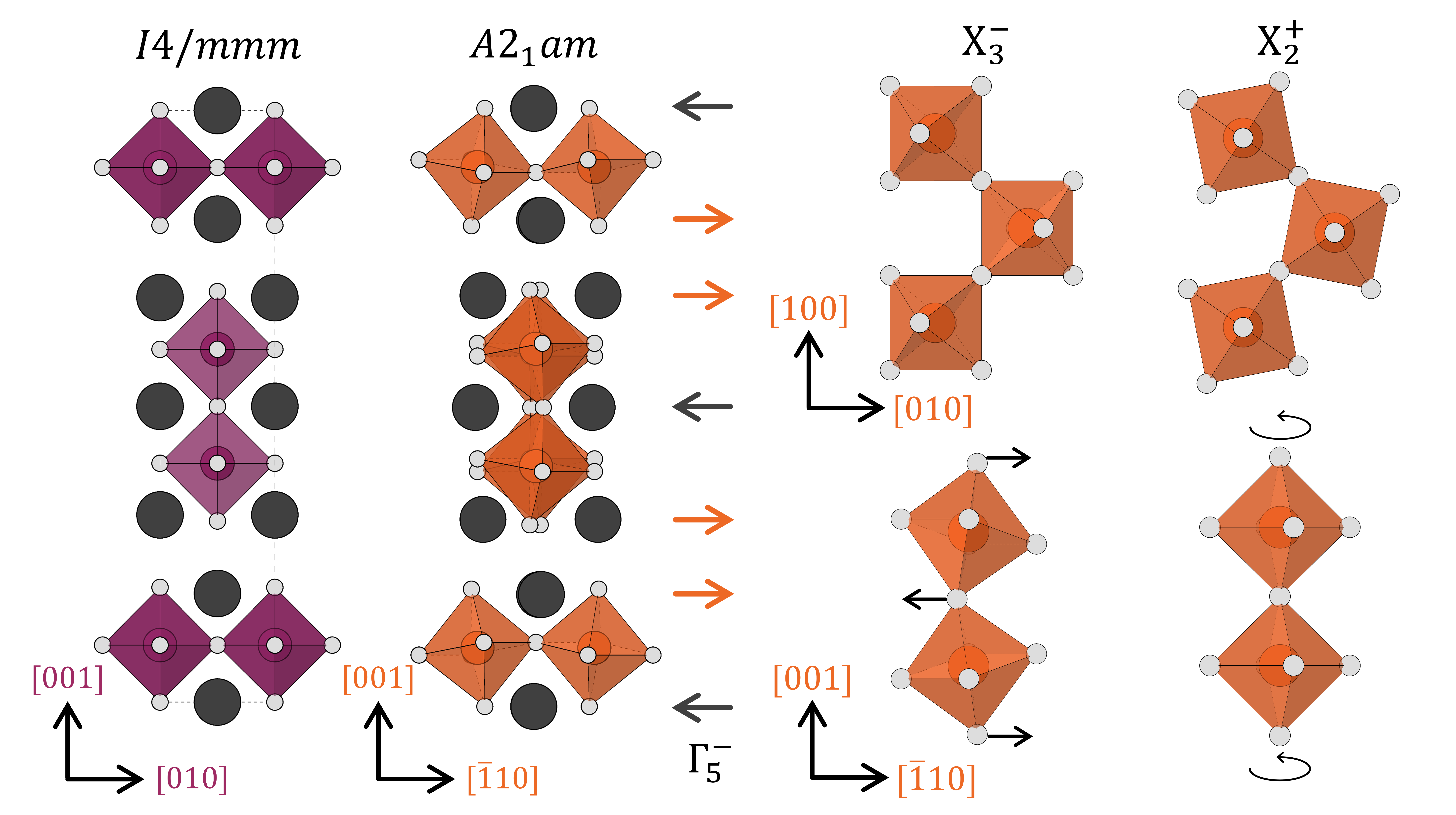}
\caption{Crystal structure of Ca$_{2.15}$Sr$_{0.85}$Ti$_{2}$O$_{7}$, showing the $I4/mmm$ aristotype and the polar $A2_1am$ structure resulting from the TiO$_6$ octahedral rotations (X$_2^+$) and tilts (X$_3^-$). Arrows indicate the polar displacements associated with the $\Gamma_5^-$ mode. $A$-site (Ca/Sr) cations are shown in dark grey.}
\label{structure}
\end{figure}

Hybrid improper ferroelectrics of the Ruddlesden-Popper family provide a well-defined platform for studying the interplay of multiple coupled order parameters. In these systems, polarisation emerges as a secondary order parameter through a trilinear coupling between two non-polar distortions and a polar mode. A representative example is Ca$_{2.15}$Sr$_{0.85}$Ti$_{2}$O$_{7}$ (CST), where the polar $A2_1am$ structure arises due to distortions of the tetragonal $I4/mmm$ aristotype, as illustrated in Fig.~\ref{structure}. These include an out-of-phase tilt about [110], transforming as irreducible representation X$_{3}^{-}(a;0)$ and an in-phase rotation of the TiO$_6$ octahedra about [001], transforming as X$_{2}^{+}(b;0)$. These two non-polar distortions are trilinearly coupled to a polar mode, $\Gamma_5^-$, which drives a two-against-one displacement of the Ca/Sr cations and results in a net polarisation. 

The coupling of X$_2^+$ and X$_3^+$ generates a large number of symmetry related domain variants, giving rise to complex domain wall structures. Experimental studies on Ca$_{3-x}$Sr$_{x}$Ti$_{2}$O$_{7}$ have revealed this diversity: piezoresponse force microscopy (PFM) and transmission electron microscopy (TEM) for $x=0.54$ has shown a variety of charged domain walls \cite{oh_experimental_2015}, while TEM imaging at $x=0.45$ revealed vortex-like domain configurations \cite{huang_domain_2016}. Theoretical modelling and electron microscopy indicate that walls for $x=0.5$ have predominantly N\'eel-like character \cite{lee_hidden_2017}, while in $x=0$, infrared nano-spectroscopy has revealed ferroelastic walls 60 -- 100 nm wide, attributed to gradual rotations of the X$_{2}^{+}$ and X$_{3}^{-}$ order parameters \cite{smith_infrared_2019}. 

The same X$_2^+$ rotation and X$_3^-$ tilt modes that define the domain structure also govern the sequence of structural phase transitions in CST, $A2_1am \rightarrow Amam \rightarrow Pnnm \rightarrow P4_2/mnm \rightarrow I4/mmm$ \cite{kratochvilova_mapping_2019, pomiro_first-_2020}. On heating, the transition is driven initially by a continuous reduction in the amplitude of the X$_{2}^{+}$ rotation. This is followed by a rotation of the X$_{3}^{-}$ tilt in order parameter space and reduction in magnitude as the structure approaches the $I4/mmm$ aristotype. For comparison, in Ca$_3$Ti$_2$O$_7$, the amplitudes of X$_{2}^{+}$ and X$_{3}^{-}$ evolve in an approximately linear relationship with respect to one another on heating between 100 and 500 K, as determined from X-ray powder diffraction \cite{senn_negative_2015}.

To directly investigate these distortions, we employ scanning three-dimensional X-ray diffraction (s3DXRD). This technique combines bulk sensitivity with high spatial resolution, enabling the mapping of structural distortions and their evolution over extended length scales. In contrast to commonly used methods for imaging ferroic structures, such as TEM and PFM, s3DXRD is non-destructive and provides true bulk sensitivity, thereby avoiding surface-related and preparation-induced effects. It grants access to full crystallographic information by recording the complete diffraction signal, rather than being limited to selected reflections.

In this approach, a focused X-ray beam is centred on the sample while it is rotated, and diffraction patterns are collected on an area detector at discrete angular intervals. The sample is then translated incrementally, and the measurement is repeated across the entire region of interest. The beam size is selected to be comparable to or smaller than the characteristic features under investigation. In principle, this method closely resembles conventional single-crystal X-ray diffraction. In the resulting dataset, each reflection is described by its scattering angle (2$\theta$) and azimuthal angle ($\eta$), and is correlated to its spatial origin within the sample through the rotation angle ($\omega$) and sample translation ($y$). The inclusion of spatial translation as an additional dimension distinguishes s3DXRD from conventional single-crystal diffraction, in which only angular information is recorded.

In this work, we use scanning three-dimensional X-ray diffraction to map the spatial variation of coupled order parameters in Ca$_{2.15}$Sr$_{0.85}$Ti$_2$O$_7$ and reveal the microscopic structure of its ferroelastic domain walls. We track the X$_2^+$ and X$_3^-$ modes across the walls to uncover the mechanism by which they evolve. More broadly, this work demonstrates the potential of s3DXRD as a powerful tool for probing complex spatial order in functional oxides, providing insight into the fundamental mechanisms governing domain wall formation and properties.

\section{Methods}
\subsection{Sample preparation}
Polycrystalline samples of CST were prepared by standard solid state methods, with the procedure described by Clarke \textit{et al.} \cite{clarke_situ_2021}. From this batch, a single crystal of approximate dimensions 15~$\times$~7~$\times$~35 \textmu m was selected and secured on a MiTeGen cryoloop using a UV-curing resin. Care was taken to ensure that only a single crystallite was mounted and the presence of orthorhombic twinning was confirmed by polarised light microscopy.

\subsection{Scanning 3DXRD measurements}
s3DXRD measurements were performed in Experimental Hutch 3 at the ID11 beamline at the European Synchrotron Radiation Facility, using a 43.5 keV beam focused to approximately 120 nm in diameter. The associated datasets are available \cite{ESRF_5185, ESRF_5590}. Sinogram scans were taken as the sample was alternately rotated forwards and backwards over 180$^\circ$ with a step size of 0.05$^\circ$. The diffractometer was translated with a step size of 50 nm over a 21 \textmu m range across the full width of the sample. 2D diffraction patterns were collected on a DECTRIS Eiger2 X CdTe 4M area detector, positioned approximately 150 mm downstream of the sample. Measurements were performed at 300 K. 

\subsection{Tomographic image reconstruction}
Diffraction data were processed using the ImageD11 software package \cite{wright_fable-3dxrdimaged11_2025}. Individual diffraction spots were identified, corrected \cite{wright_using_2022} and converted to reciprocal space coordinates using instrument calibration from a silicon single crystal. Reflections were indexed on a 5.44~$\times$~5.46~$\times$~19.69~\AA~unit cell and ($hkl$) indices were assigned. Reconstruction of the crystal shape proceeded using tomographic back projection of the peak intensities versus diffractometer $y$-coordinate. The reconstruction was enhanced using the maximum likelihood expectation maximisation (MLEM) algorithm \cite{bonnin_impurity_2014}. 

Using the full diffraction signal, the overall shape of the crystal can be reconstructed. The resulting image is a slice through the centre of the crystal, where the constrast is given by diffraction intensity. Alternatively, it is possible to reconstruct images using a meaningful subset of reflections, enabling selective visualization of specific structural features. In the case of CST, there are two orthorhombic twin/ferroelastic domains related by a fourfold rotation about the $c$-axis. These domains can be distinguished based on their reflection conditions: $k + l = \mathrm{even}$ for the $A$-centred domain, and $h + l = \mathrm{even}$ for the $B$-centred domain. This distinction arises from the orientation of the octahedral tilting. Reflections where $h,k,l =$ even are excluded from this filtering procedure.

\subsection{Symmetry-adapted strain analysis} \label{lp_extraction}
The far-field geometry employed in s3DXRD, where the detector is positioned at a large distance from the sample, allows for precise measurement of diffraction peak positions. This sensitivity to small shifts in peak positions enables detection of lattice parameter perturbations with high precision, achieving strain resolutions on the order of $10^{-4}$ \cite{borbely_calibration_2014}. 

Local variations in the lattice parameters were extracted using a tomographic refinement algorithm adapted from the method of Henningsson and Hall \cite{henningsson_efficient_2023}, and implemented via the ASTRA toolbox \cite{van_aarle_fast_2016}. Each scan
position provides a projection of the local lattice parameters along the beam direction. By combining projections collected at multiple positions and angles, a full three-dimensional map of the lattice parameters can be reconstructed tomographically. This approach fits a global model to all voxels simultaneously, accounting for the contribution of each voxel to the integrated diffraction signal. 

First, the crystal shape was reconstructed. Then, for each voxel, the local lattice parameters were refined by fitting the observed peak positions to the metric tensor equation:
\begin{equation}
    \frac{1}{d^{2}}=Ah^{2}+Bk^{2}+Cl^{2}+Dkl+Ehl+Fhk
\end{equation}

An initially uniform 2D image of the metric tensor elements ($A-F$) was forward projected to generate a sinogram by weighting each of the six layers according to the grain shape reconstruction. The resulting $1/d^2$ values for each $hkl$ projection in the 2D sinogram were then compared to observed peak positions. The fit was optimised by minimising the sum of squares of the differences between the observed and calculated $1/d^2$ values using a least-squares optimisation method (L-BFGS algorithm in SciPy). This refinement procedure applied no symmetry constraints or regularisation, enabling an unbiased reconstruction of local distortions. Unlike conventional strain tensor approaches, this method does not require an explicit reference `d-zero' lattice, as the refinement directly fits the metric tensor elements.

These basis vectors were then used to compute symmetry-adapted modes with respect to the tetragonal $I4/mmm$ aristotype structure. In principle, the lattice parameters of the aristotype would provide the natural reference, but this phase occurs at temperatures well above those accessible in our s3DXRD measurements and could not be measured directly. Instead, reference values of $a$ and $c$ were defined as the average lattice parameters across all voxels and $\alpha = \beta = \gamma$ were fixed to 90$^\circ$.

\begin{figure}[h]
\includegraphics[width=0.48\textwidth]{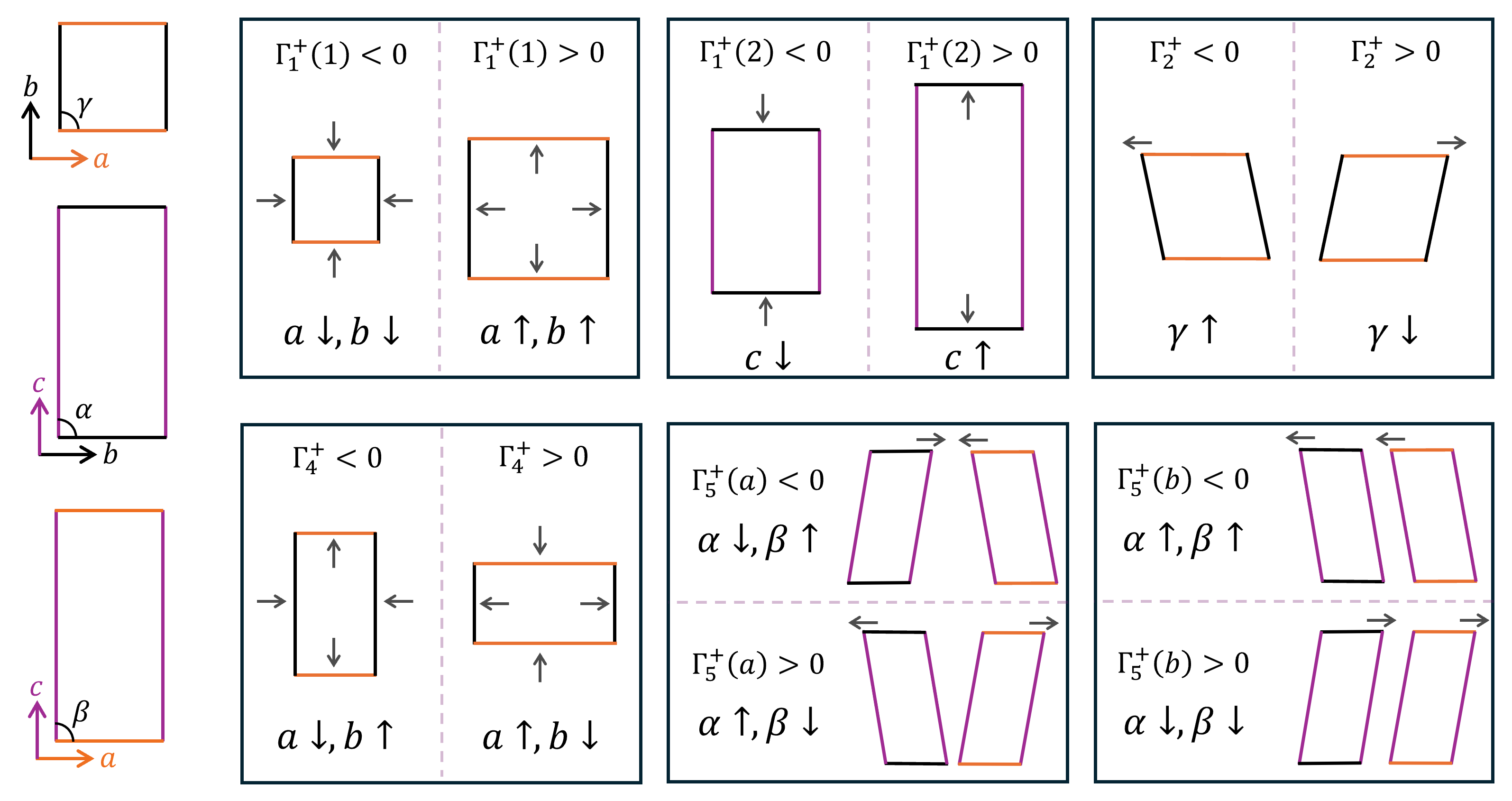}
\caption{Schematic depiction of the symmetry-adapted strain modes, labelled relative to the irreducible representation of the $I4/mmm$ aristotype \cite{stokes_h_t_isodistort_nodate}, but shown relative to the basis vectors of the $F4/mmm$ setting.}
\label{strain_schem}
\end{figure}

To achieve this, the refined lattice basis vectors were first expressed in Cartesian coordinates and compared to the basis vectors of the hypothetical parent cell - adopted in the $F4/mmm$ setting to avoid a later strain transformation to the distorted setting - to obtain unit-less parent cell strains. The symmetry-adapted strain modes could then be computed. Full details provides in the Supplemental Material \cite{Supplemental_Material}. These modes are defined relative to the $I4/mmm$ aristotype and have clear physical interpretations, demonstrated schematically in Fig.~\ref{strain_schem}: $\Gamma_{1}^{+}(1)$ corresponds to a symmetric expansion/contraction within the $ab$ plane that preserves tetragonal symmetry, $\Gamma_{1}^{+}(2)$ is a uniform expansion/contraction of the $c$ axis, $\Gamma_{2}^{+}$ is a shear strain within the $ab$ plane, $\Gamma_{4}^{+}$ is the orthorhombic strain and corresponds to an antisymmetric distortion between the $a$ and $b$ axes, $\Gamma_{5}^{+}(a)$ and $\Gamma_{5}^{+}(b)$ describe shear strains between the $ab$ plane and $c$ axis. 

The strain mode decomposition was carried out in ISODISTORT \cite{stokes_h_t_isodistort_nodate, campbell_b_j_isodisplace_2006} using Method 3 with space group symmetry $P1$ and $F4/mmm$ basis ((1, 1, 0), (-1, 1, 0), (0, 0, 1)). The resulting mode definitions were exported in \texttt{TOPAS.STR} format, from which parent-cell and symmetry-adapted strain modes were derived and subsequently implemented in a custom Python script.

\subsection{Full symmetry-mode refinement} \label{sm_extraction}
To gain a complete picture of the local structural distortions, it would be beneficial to perform a full symmetry-adapted structural refinement in each voxel. This approach would allow for the direct extraction of the amplitude and phase of the X$_{2}^{+}$, X$_{3}^{-}$, and $\Gamma_5^-$ modes. However, implementing such an analysis within the tomographic framework described above presents a considerable technical and computational challenge. Consequently, an alternative approach has been adopted to obtain preliminary results. 

In this method, a small region of the sample is selected, corresponding to a path through the sinogram. The region must be sufficiently large to include an adequate number of reflections for reliable refinement. These reflections are then extracted and written to a \texttt{.hkl} file, which forms the basis for subsequent structural analysis. Refinements were then performed in TOPAS 7 \cite{coelho_topas_2018} using the symmetry-mode formalism, to refine the relevant symmetry modes. This approach parametrises structural distortions directly in terms of the mode amplitudes and phases. By repeating this process systematically across the sample, a spatial map of the amplitude and phase of the X$_{2}^{+}$, X$_{3}^{-}$, and $\Gamma_5^-$ can be constructed. 

A key limitation of this method is the assumption that each reflection originates from a single point within the sample. In practice, however, the diffracted signal arises from the entire illuminated volume defined by the intersection of the incident beam with the sample \cite{henningsson_reconstructing_2020, henningsson_efficient_2023}. As a result, each measured reflection represents an integrated signal over this volume, effectively averaging the structural information along the beam path. Consequently, the refined parameters associated with a given voxel are influenced by all illuminated voxels at that scan position, leading to reduced spatial resolution and accuracy.

The data presented using this method were obtained from the same sample, but using a different crystallite and during a different experiment (beamtime HC-5185) from that used for the reflection-condition reconstruction and symmetry-adapted strain analysis (beamtime HC-5590). Further details are provided in the Supplemental Material.

\section{Results}
Fig.~\ref{twinrecon} shows a slice through the crystal, reconstructed tomographically both from the full diffraction signal and from a subset of reflections satisfying the $A$- and $B$-face centering conditions. This reveals an interlocking ferroelastic domain pattern, with domain walls orientated parallel to [001] and perpendicular to [110] and [1$\bar{1}$0]. The domains exhibit a tapered, roughly triangular shape, which is consistent with a slight misorientation of the [001] axis relative to the image plane.

\begin{figure}[h]
\includegraphics[width=0.48\textwidth]{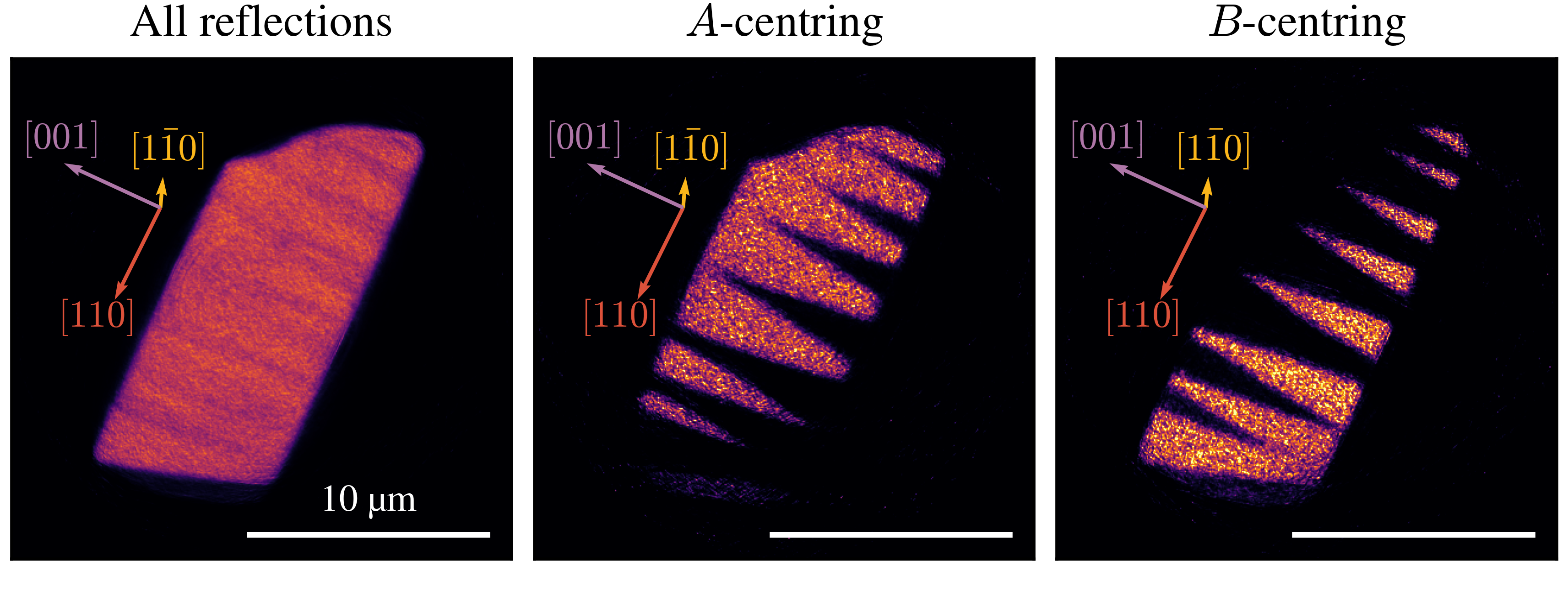}
\caption{Tomographic reconstruction using all reflections and reconstruction of orthorhombic twin/ferroelastic domains based on the $A$- and $B$-centring reflection conditions.}
\label{twinrecon}
\end{figure}

\begin{figure}[h]
\includegraphics[width=0.48\textwidth]{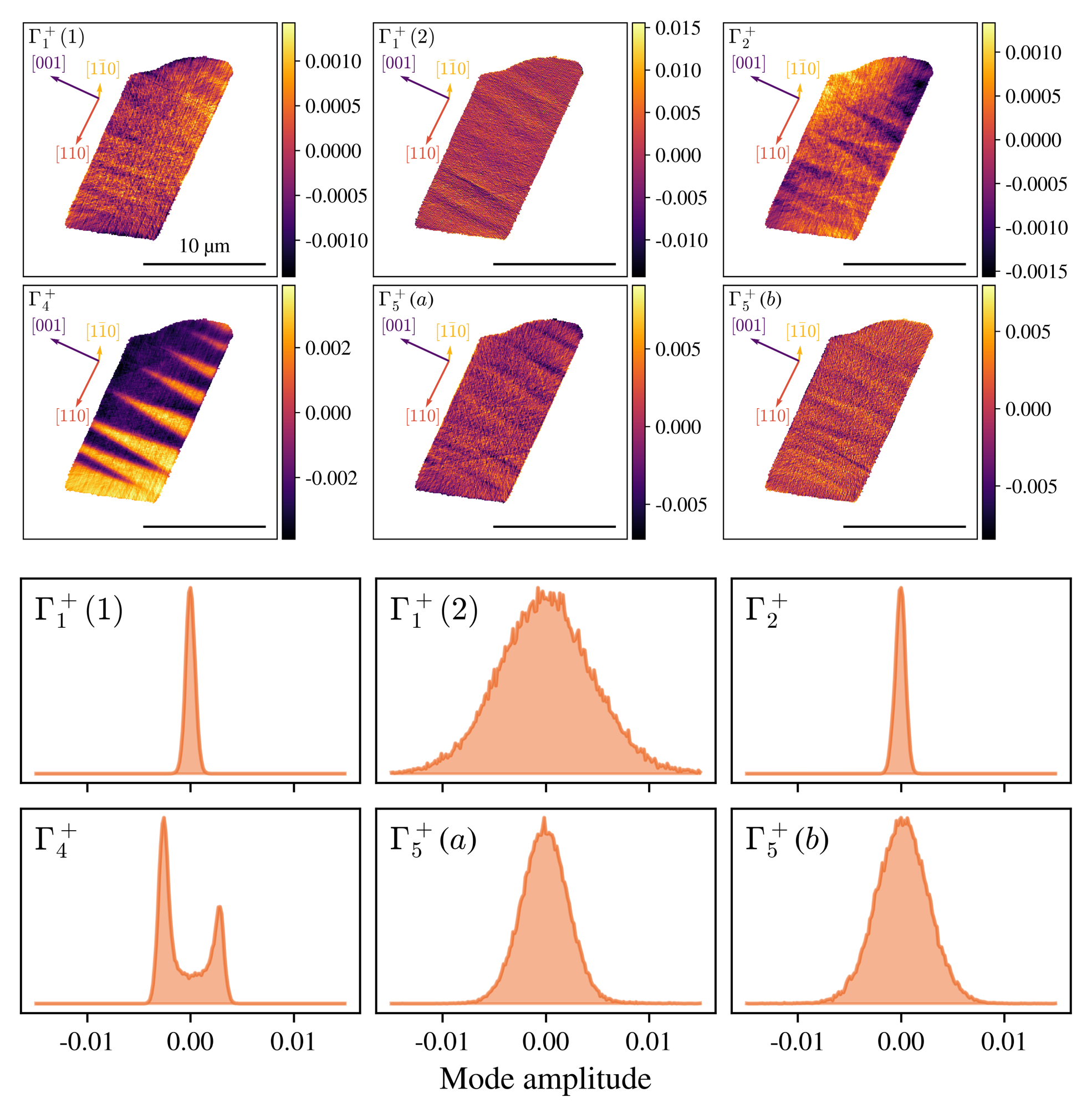}
\caption{Spatially resolved maps and associated histograms of symmetry-adapted strain modes. In the $\Gamma_4^+$ histogram, the two peaks correspond to the two twin domains, while the intensity at $\Gamma_4^+ = 0$ is consistent with extended domain-wall centres. Note that different scale ranges are used for each mode map to account for variations in strain magnitude.}
\label{strains}
\end{figure}

To examine the structural distortions underlying this domain pattern, symmetry-adapted strain modes were calculated relative to the $I4/mmm$ aristotype and mapped spatially, as shown in Fig.~\ref{strains}. Among these, the $\Gamma_4^+$ mode exhibits a pronounced spatial variation in a stripe-like pattern where the strain alternates between neighbouring twin domains, mirroring the reconstruction of the reflection conditions (Fig.~\ref{twinrecon}). Acting as a secondary order parameter to the octahedral distortions X$_2^+$ and X$_3^-$, $\Gamma_4^+$ captures the ferroelastic domain structure. Owing to its superior strain resolution compared with systematic intensity errors, our analysis primarily focuses on the strain reconstructions. The remaining modes show weaker, more diffuse variations. The $\Gamma_2^+$ mode broadly follows the domain morphology with reduced amplitude, consistent with elastic compatibility at the domain walls, while $\Gamma_5^+(a)$ exhibits only subtle, localised modulations near the walls, indicating additional strain accommodation.

Statistical analysis of the strain distributions further supports these observations. Histograms of all modes are centred near zero, with the notable exception of $\Gamma_4^+$, which exhibits a clear bimodal distribution. The two peaks correspond to the oppositely signed orthorhombic distortions in the twin variants, while a finite intensity at $\Gamma_4^+ = 0$ is consistent with extended wall centres. An additional feature is the relatively broad distribution of the $\Gamma_1^+(2)$ mode, corresponding to uniform strain along the $c$-axis. This broadening likely reflects spatial variations in the $c$ lattice parameter arising from structural inhomogeneities such as Ruddlesden–Popper intergrowths or stacking faults. The spatial maps of $\Gamma_1^+(2)$ also reveal slight modulations at the domain walls, consistent with elastic compatibility constraints.

\begin{figure}[h]
\includegraphics[width=0.48\textwidth]{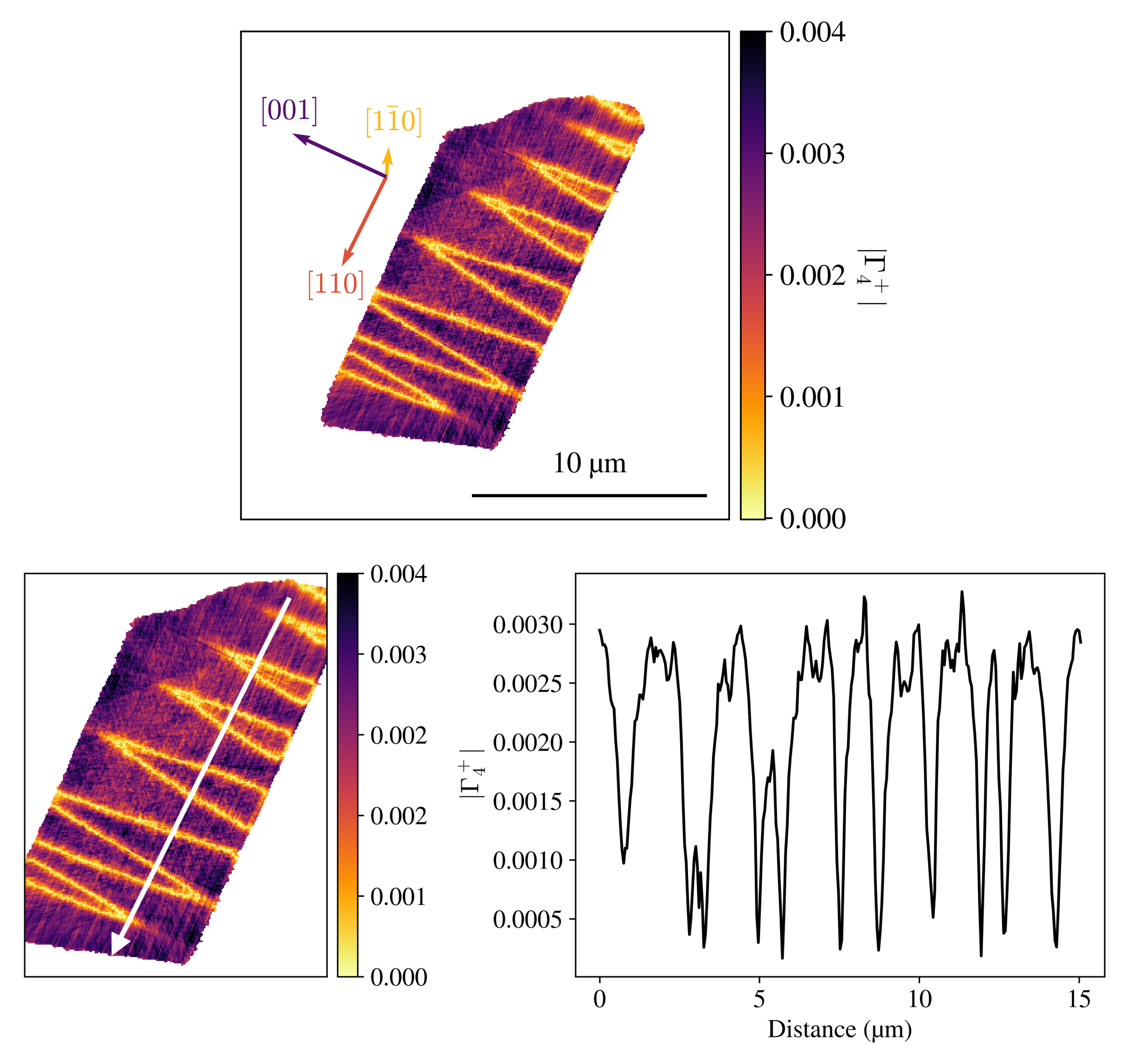}
\caption{Spatially resolved map of the absolute strain magnitude $|\Gamma_4^+|$ with line profiles taken across multiple domains, revealing the presence of broad, tetragonal-like domain walls.}
\label{modgamma4}
\end{figure}

To further quantify the domain wall structure, the magnitude of the orthorhombic strain, $|\Gamma_4^+|$, was analysed (Fig.~\ref{modgamma4}). This reveals extended domain wall–like regions in which the orthorhombic strain is significantly suppressed, approaching zero. Line profile analysis indicates that these regions have a characteristic width of approximately 120 nm, which is notably large for ferroelastic domain walls.

\begin{figure}[h]
\includegraphics[width=0.48\textwidth]{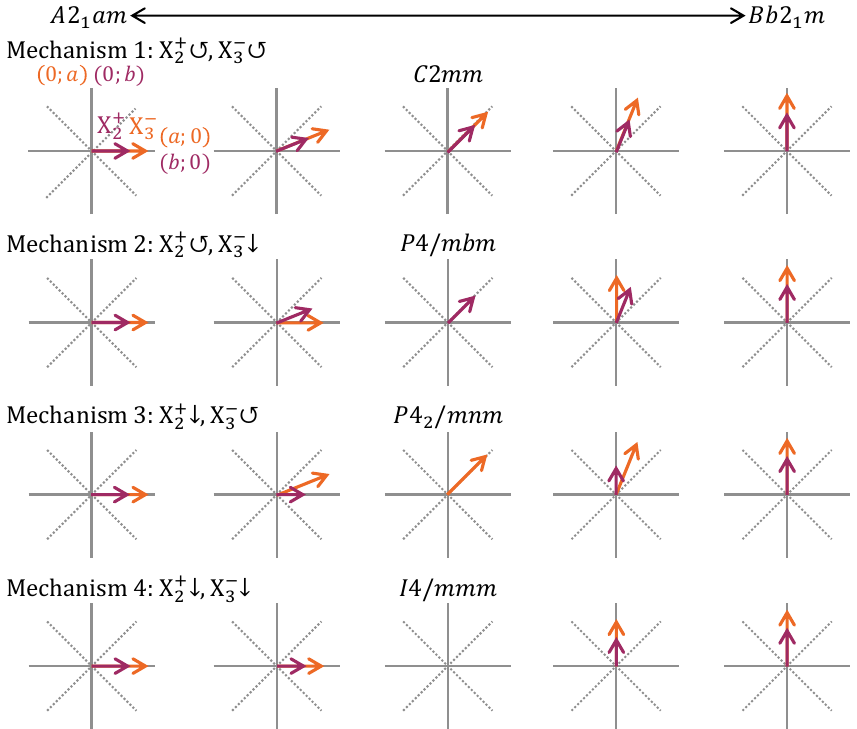}
\caption{Switching pathways of ferroelastic domain walls from $A2_1am$ to $Bb2_1m$, revealing four mechanisms where the $X_{3}^{-}$ and $X_{2}^{+}$ modes either rotate within order-parameter space or diminish to zero.}
\label{mechanism}
\end{figure}

Local minima in $|\Gamma_4^+|$ coincide with the centres of the domain walls, indicating a recovery of near-tetragonal symmetry in these regions. As the orthorhombic distortion arises from the coupled octahedral tilt and rotation modes (X$_2^+$ and X$_3^-$), this implies that at least one of the primary order parameters must be either suppressed or reorientated at the wall. This places direct constraints on the microscopic pathway connecting neighbouring domains. Four mechanisms can be envisaged, as illustrated in Fig.~\ref{mechanism}:
\begin{enumerate}
    \item X$_{3}^{-}$ and X$_{2}^{+}$ rotate to $(a;a)$ and $(b;b)$, respectively, giving an intermediate $C2mm$;
    \item X$_{3}^{-}$ decreases to $(0;0)$ while X$_{2}^{+}$ rotates to $(b;b)$, giving an intermediate $P4/mbm$;
    \item X$_{2}^{+}$ decreases to $(0;0)$ while X$_{3}^{-}$ rotates to $(a;a)$, giving an intermediate $P4_2/mnm$;
    \item X$_{3}^{-}$ and X$_{2}^{+}$ both decrease to $(0;0)$, recovering the aristotype $I4/mmm$.
\end{enumerate}

Mechanism 1 appears somewhat unlikely as the intermediate $C2mm$ structure is orthorhombic and therefore incompatible with the tetragonal symmetry observed experimentally at the walls. This discrepancy could, however, be reconciled if the orthorhombic strain in the $C2mm$ phase were sufficiently small to be experimentally indistinguishable. While the remaining three pathways, all involving tetragonal intermediates, cannot be unambiguously distinguished, mechanism 3 is most consistent with the available evidence. \textit{In situ} polarisation switching studies on CST, combining X-ray powder diffraction with simulations, indicate that switching pathways involving suppression of the X$_{2}^{+}$ mode are favoured \cite{clarke_situ_2021}. This is analogous to the temperature-driven phase transitions, where the evolution from $A2_1am$ to $I4/mmm$ initially proceeds by melting of X$_{2}^{+}$ \cite{pomiro_first-_2020}. These results would suggest mechanism 3 represents the low energy pathway for switching between domains. Density functional theory calculations for Ca$_3$Ti$_2$O$_7$ further support this interpretation, showing that suppression of X$_2^+$ is energetically less costly than suppression of X$_3^-$ \cite{nowadnick_domains_2016}. Although a cooperative rotation of both modes (mechanism 1, $C2mm$ intermediate) is predicted to be slightly lower in energy (41 meV/Ti) than the $P4_2/mnm$ intermediate associated with mechanism 3 (45 meV/Ti), the latter remains competitive. By contrast, the $P4/mbm$ pathway (mechanism 2) is significantly higher in energy (172 meV/Ti) and is therefore unlikely. It should be noted that these calculations do not explicitly account for the strain field at the domain wall, which may modify the relative energetics and render the $P4_2/mnm$ pathway more favourable. This balance may also change with Sr concentration. First-principles calculations suggest that the relative stability of $P4_2/mnm$ compared to ground state $A2_1am$ increases significantly from $x=0$ to $x=0.85$ \cite{pomiro_first-_2020}, with $P4_2/mnm$ predicted to become the ground state for compositions beyond $x \approx 1.25$. This trend is consistent with the experimental observation of a $P4_2/mnm$ phase at ambient conditions for $x=1$ \cite{huang_topological_2016}.

A similar conclusion has also been reached in a first-principles study of the structurally related Aurivillius ferroelectrics SrBi$_2$(Ta,Nb)$_2$O$_9$, in which ferroelastic domain walls (corresponding to 90$^\circ$ ferroelectric walls, as in CST) were predicted to favour a pathway involving a $P4_2/mnm$ intermediate \cite{pokhrel_ferroelectric_2023}. The comparison is instructive, as domain structures closely resembling those observed in CST have been reported in these systems \cite{su_different_2003, ding_ferroelectric_2001}. This supports a common idea in which the X$_2^+$ mode is suppressed while the X$_3^-$ mode undergoes reorientation across the domain wall.

\begin{figure}[h]
\includegraphics[width=0.48\textwidth]{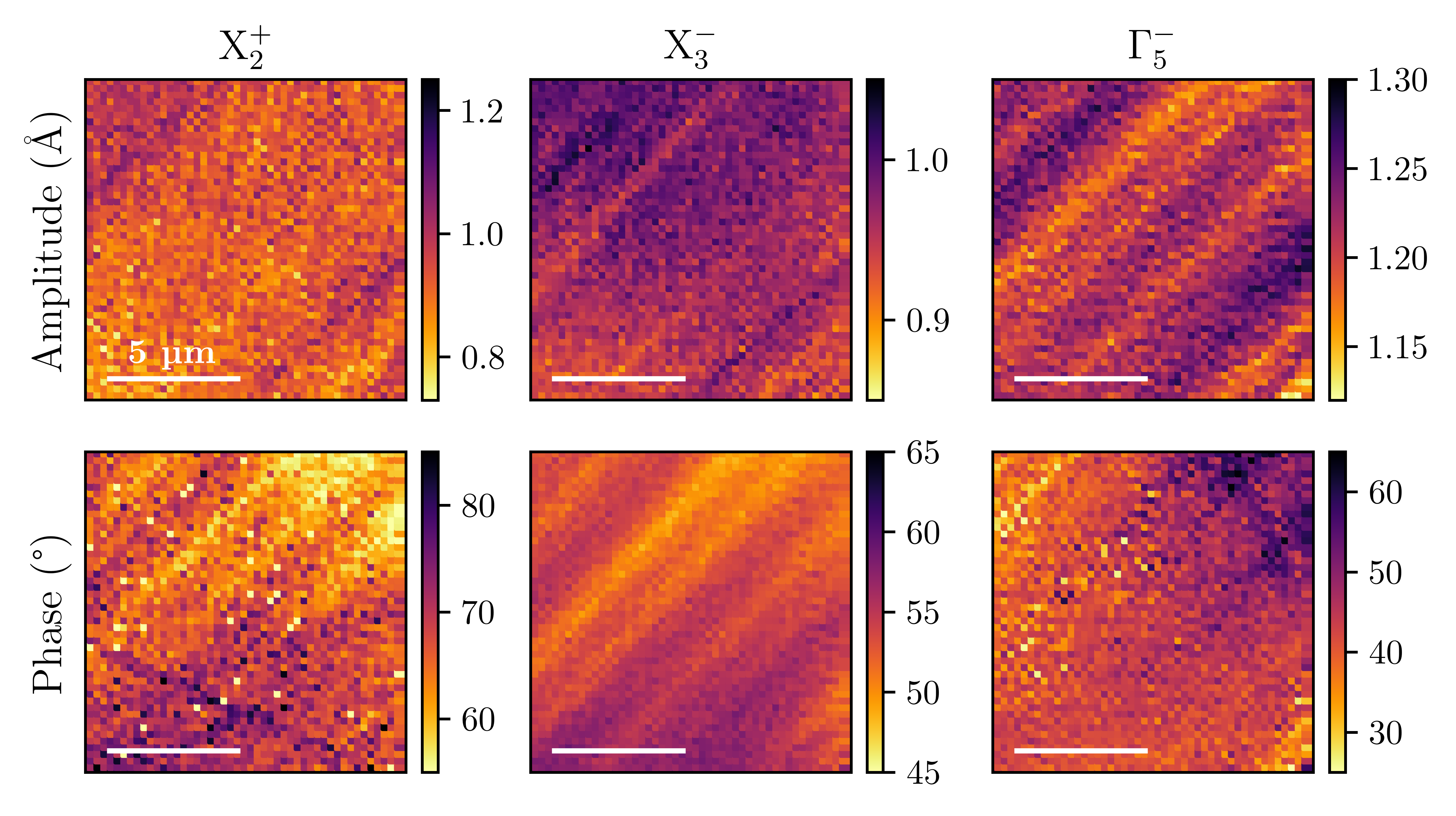}
\caption{Spatial maps of the amplitude and phase of the $X_2^+$, $X_3^-$, and $\Gamma_5^-$ modes.}
\label{modemaps}
\end{figure}

As strain maps alone cannot unambiguously resolve the symmetry at the domain wall, it is useful to examine the X$_2^+$, X$_3^-$ and $\Gamma_5^-$ modes directly. Preliminary maps of the amplitude and phase of these modes are presented in Fig.~\ref{modemaps}. As this method involves averaging over multiple projection directions and requires sufficiently large sampling volumes to ensure an adequate number of reflections for reliable refinement, true local minima at the domain wall centre are not fully resolved, and the effective spatial resolution is lower than that achieved in the symmetry-adapted strain analysis. However, fortuitously, due to the striped nature of the identified domain structure, there are a significant number of projections that do not average across different domains (i.e. the projections parallel to the stripes). So while these limitations reduce sensitivity to narrow wall features and average out some of the contrast between different order parameter directions and magnitudes, valuable insight can still be gained.

The $\Gamma_5^-$ amplitude exhibits a stripe-like modulation. It is expected that the amplitude of $\Gamma_5^-$ should diminish at the centre of ferroelastic walls \cite{nowadnick_domains_2016}, and the shallow minima observed here are consistent with this expectation once the averaging inherent to this method is considered. Ideal 90$^\circ$ phase rotations of the three modes across ferroelastic walls are not cleanly resolved, again consistent with spatial averaging. The X$_3^-$ mode displays a clear banded structure that aligns with regions of reduced $\Gamma_5^-$ amplitude. Its phase varies smoothly in stripe-like patterns across the crystal, while the amplitude exhibits a more modest variation and does not approach zero. This behaviour is indicative of a continuous rotation of the octahedral tilt through the wall, rather than simple amplitude suppression, which would produce two discrete domains separated by a sharp discontinuity. Such observations are most consistent with rotation-type mechanisms (1 or 3). While discriminating between these requires detailed analysis of X$_2^+$, given the observation of the rotation of the X$_3^-$ phase and tetragonal symmetry at the domain wall, we conclude that a suppression of the X$_2^+$ amplitude is most likely to occur. Overall, these preliminary mode maps are suggestive but not conclusive; higher-resolution mode-resolved mapping and/or complementary techniques will be required to establish the wall mechanism unambiguously.

Given the large number of possible domain variants and the additional complexity introduced by different polarisation states, it is likely that multiple microscopic pathways coexist in CST. While the present results most strongly support mechanism 3, mechanisms 2–4 all involve the emergence of local tetragonal symmetry at the wall centre. Consistent with this, Huang \textit{et al.} \cite{huang_domain_2016} indicate two types of ferroelastic walls for $x=0.45$: one retaining orthorhombic character and another exhibiting tetragonal symmetry, corresponding to mechanisms 1 and 3, respectively. It is therefore notable that only walls with tetragonal character are observed in the present study, which may reflect differences in Sr concentration.

When compared with the wall widths reported by Smith \textit{et al.} \cite{smith_infrared_2019} in Ca$_3$Ti$_2$O$_7$, the average widths observed here are somewhat larger. This difference may simply reflect the experimental technique, since infrared nano-spectroscopy is surface-sensitive, whereas s3DXRD probes the bulk. Alternatively, it may arise from differences in the microscopic mechanisms. Smith \textit{et al.} report finite mode amplitudes for both X$_2^+$ and X$_3^-$ across the wall, consistent with mechanism 1. More generally, it has been shown that symmetry constraints requiring an octahedral rotation or tilt amplitude to vanish at the wall centre tend to increase the domain wall thickness \cite{cao_landau-ginzburg_1990, xue_orientations_2014}. This provides a natural explanation for the comparatively large wall widths observed in CST. Even in the case of a $C2mm$ intermediate, X$_2^+$ is suppressed in alternating perovskite slabs, while in the scenario favoured here (mechanism 3), the complete suppression of X$_2^+$ would further promote broadening of the wall. This may account for the increased thickness relative to Ca$_3$Ti$_2$O$_7$.

We have recently utilised the s3DXRD approach to investigate the ferroelastic domain structure in the orthorhombic phase of $n=1$ Ruddlesden-Popper La$_{1.675}$Eu$_{0.2}$Sr$_{0.125}$CuO$_4$.\cite{ladbrook2025giant} This revealed a strikingly similar domain morphology to that observed in CST, characterised by extended domain walls despite the markedly different electronic properties. In that system, the relevant octahedral tilt mode (X$_3^+$) evolves through a continuous rotation in order-parameter space. The emergence of wide domain walls in both materials, driven by octahedral tilting, suggests that such structural heterogeneity is a generic feature of Ruddlesden–Popper phases. This points to common underlying principles governing domain wall formation, independent of chemistry or functionality. This perspective motivates extending such studies to other Ruddlesden–Popper systems exhibiting complex tilt-driven structural transitions, such as superconducting Sr$_2$RuO$_4$ and La$_3$Ni$_2$O$_7$, as well as magnetoelectric Ca$_3$(Mn,Ti)$_2$O$_7$. In the latter, unusual curved ferroelastic domain structures have previously been reported \cite{huang_domain_2016}, potentially arising from the coexistence of order parameters that drive distinct phase transitions to the improper hybrid ferroelectric phase $A2_1am$ and the uniaxial negative thermal expansion phase $Acaa$ \cite{senn_negative_2015}. Establishing whether similar mechanisms govern domain-wall formation across these systems would provide valuable insight into the universality of tilt-mediated structural heterogeneity in layered oxides.

\section{Conclusion}
The combination of tomographic reconstruction and symmetry-adapted strain mapping in CST reveals that ferroelastic domain walls are intrinsically extended and structurally coherent. The results show that wall formation is governed by a continuous rotation of the X$_3^-$ tilt mode accompanied by a suppression of the X$_2^+$ rotation, resulting in near-tetragonal symmetry at the wall centre. These findings establish that the interplay of coupled order parameters dictates domain-wall width, orientation, and local strain accommodation, thereby directly shaping the ferroic response.

More broadly, this work highlights the emerging capability of s3DXRD to resolve atomic-scale distortions to mesoscale domain structures.\cite{ladbrook2025giant,SimpsonPRL2026}, By providing direct access to the structural mechanisms governing ferroelastic domain walls, this approach opens new opportunities to understand, engineer, and ultimately exploit functionalities at domain boundaries in complex oxides.

\begin{acknowledgments}
E.L. thanks the University of Warwick for a PhD studentship through the Warwick Centre for Doctoral Training in Analytical Science. M. S. S. acknowledges the Royal Society for a fellowship (UF160265 $\&$ URF$\backslash$R$\backslash$231012) for funding.  Initial sample characterisation was performed using equipment provided
by the Warwick X-Ray Diffraction Research Technology Platform.  We acknowledge the European Synchrotron Radiation Facility (ESRF) for provision of beamtime at ID11 under proposal numbers HC-5590 \cite{ESRF_5590} and HC-5185 \cite{ESRF_5185}.
\end{acknowledgments}

\bibliography{apssamp}

\clearpage
\newpage

\onecolumngrid


\end{document}